\documentclass[11pt]{article}

\usepackage{geometry}
\usepackage[utf8]{inputenc} 
\usepackage[T1]{fontenc}    
\usepackage{hyperref}       
\usepackage{url}            
\usepackage{booktabs}       
\usepackage{amsfonts}       
\usepackage{nicefrac}       
\usepackage[activate={true,nocompatibility},final,tracking=true,kerning=false,spacing=false,factor=1100,stretch=10,shrink=10]{microtype}            
\usepackage[normalem]{ulem} 
\usepackage{lipsum}
\usepackage{color}
\usepackage{amsmath}
\usepackage{graphicx}
\usepackage{subcaption}
\usepackage{titlesec}
\usepackage{fourier}
\usepackage{times}
\usepackage[colorinlistoftodos]{todonotes}
\usepackage{ulem} 
\usepackage[ruled,vlined]{algorithm2e} 
\usepackage{authblk}
\usepackage{cite}           
\usepackage{epstopdf}

\usepackage{caption}
\definecolor{webgreen}{rgb}{0,.35,0}
\newcommand{\jpb}[2]{\textcolor{webgreen}{\textit{\sout{#1}#2}}}

\title{Accelerating GMRES with Deep Learning in Real-Time}

\author[1,3]{Kevin Luna \thanks{kevinluna@math.arizona.edu}}
\author[2,3]{Katherine Klymko\thanks{kklymko@lbl.gov}}
\author[3]{Johannes P. Blaschke \thanks{jpblaschke@lbl.gov}}

\affil[1]{Program in Applied Mathematics, The University of Arizona, USA}
\affil[2]{Computational Research Division, Lawrence Berkeley National Laboratory, USA}
\affil[3]{National Energy Research Scientific Computing Center, Lawrence Berkeley National Laboratory, USA}

\begin{document}

\date{}

\maketitle


\begin{abstract}
    GMRES is a powerful numerical solver used to find solutions to extremely
    large systems of linear equations. These systems of equations appear in
    many applications in science and engineering. Here we demonstrate a
    real-time machine learning algorithm that can be used to accelerate the
    time-to-solution for GMRES. Our framework is novel in that is integrates
    the deep learning algorithm in an \emph{in situ} fashion: the
    AI-accelerator gradually learns how to optimizes the time to solution
    without requiring user input (such as a pre-trained data set).
    
    We describe how our algorithm collects data and optimizes GMRES. We
    demonstrate our algorithm by implementing an accelerated (MLGMRES) solver
    in Python. We then use MLGMRES to accelerate a solver for the Poisson
    equation~--~a class of linear problems that appears in may applications.
    Informed by the properties of formal solutions to the Poisson equation, we
    test the performance of different neural networks. Our key takeaway is that
    networks which are capable of learning non-local relationships perform well, without
    needing to be scaled with the input problem size, making them good
    candidates for the extremely large problems encountered in high-performance
    computing. For the inputs studied, our method provides a roughly 2$\times$
    acceleration.
\end{abstract}


\section{Introduction}

The list of scientific problems that are modeled by partial differential
equations (PDEs) is vast.  A numerical solution to a PDE is therefore
incredibly useful for the study of the complex relationships arising in the
world around us.  From the study of microscopic particle interactions~\cite{cai2014} to the
study of massive astronomical objects and nearly every length scale in between,
solutions to PDEs are needed~\cite{AMReX_JOSS,zingale2019castro}.  In practice, most applications require solutions
to these PDEs as a function of time.  This often requires the numerical
computation of these PDEs at many instances called “time steps”.  At these time
steps, solvers are employed to compute solutions.  The precise details of these
solvers depends highly on the mathematical properties and scientific context of
the underlying problem being solved.  Despite this large variation in methods
used to obtain solutions to PDEs, there is often overlap in some of the details
of these methods.  For example more often than not, at some point a linear
problem will need to be solved using iterative
methods~--~i.e. the solution is refined with each iteration.  The rate
at which these methods converge to the final solution varies substantially with
the problem and tends to depend on initial information provided to the solver.
This initial information often takes the form of a problem-specific initial
guess or a problem-specific transformation (a preconditioner) that leads to more agreeable convergence properties.  Consequently, the
performance of these methods can be significantly improved by supplying such
initial conditions and preconditioners \cite{PrecondBenzi}. In practice,
finding effective initial guesses and preconditioners (if any can be found at
all using existing methods) for a particular problem is very much an art that
requires thoughtful input from an expert. In addition to accelerating
simulations through algorithmic optimizations (such as finding better choices for preconditioners), we see great potential in taking
advantage of the rapid improvements in hardware targeting machine learning
applications.

As high-performance computing paradigms shift towards ever increasing
heterogeneity, the abundance of accelerators aimed at machine learning
applications provide an increasingly attractive opportunity to make use of
these resources to accelerate traditional scientific simulations~\cite{whitelam2019learning,whitelam2019evolutionary,beeler2019optimizing,tsai2020lstm,pathak2020amrex,hutson2020scimag,kasim2020billion,pham2018efficient}. Enabled by
the presence of these accelerators in HPC systems, we envision using neural
networks to learn efficient preconditioners and initial guesses for simulations
with underlying iterative linear solvers in real-time. To demonstrate the
merits of this idea, here we develop a real-time deep learning powered
accelerator for GMRES~\cite{saad1986gmres}
applied to
second-order linear partial differential equations.
Countless applications in
heat flow, diffusion phenomena, wave propagation,  fluid flow, electrodynamics,
quantum mechanics, and much more are faithfully modeled by linear second order
PDEs. Consequently, by building a foundation on such a useful class of
problems, we are able to establish out of the gate that the underlying ideas
introduced here inherently posses the potential to be useful for a wide range
of problems of practical importance.

Broadly speaking, we aim to utilize deep learning to accelerate PDE-based
simulations by training and applying these learned preconditioners in
real-time.  We emphasize that the goal here is to accelerate a simulation in
real-time when we do not have access to relevant data prior to the start of the
simulation. In particular, we demonstrate that we can train a neural network in
real-time to learn preconditioners as a simulation runs~--~improving the speed
of the underlying linear solver and decreasing the overall wall-clock-time of the
simulation.  In shifting our focus to accelerating the solver in real-time
while data is being generated, some non-standard challenges are encountered.
Among these challenges, we must deal with the fact that we are working with new
data in a quasi-online fashion as it becomes available over the course of our
simulation. In other words, we do not necessarily (and often do not) have
access to a representative distribution of data as our simulation progresses.
Another key challenge when working in real-time lies in the selection and
training of a model.  Simply put, if our learned model is to be constructed and
used in real time to accelerate a simulation, the training time scale cannot be
comparable to the time scale of the simulation.  Owing to these challenges,
working with neural networks in real-time with limited incoming data ("online")
is difficult. We overcome these challenges with deep learning for the problem
at hand by carefully selecting a model that aims to accelerate the underlying
solver rather than accurately learn the solution i.e. we are \emph{not}
merely learning a look-up table for the solution given the inputs), and by
curating the training set as it grows in real-time so that the neural network
is able to improve performance before a representative distribution of data is
attained.

\subsection{The Case for Online Acceleration vs. Offline End-to-End Learning}

While deep learning has shown great potential in producing models that capture
complex relationships in scientific applications and even discovering such
relationships, some key underlying difficulties have persisted despite this
progress \cite{MLChallenges,MLDataProblem}.

In the space of scientific simulation and modeling, there are two such
difficulties.  The first is the ``trust problem'': how do we trust the result
of a neural network? A common approach is to treat outputs from neural networks
as probabilistic, thus treating outputs as confidence intervals. This might
work well, e.g. if the goal is to predict the preferences of most customers.
However there is little room for uncertainty when solving PDEs: a mathematical
object either satisfies the PDE~--~i.e. it is a solution~--~or it does
not.  The second  difficulty is that most machine learning workflows assume
that data is both cheaply and readily available. While this is not a bad
assumption for many commercial applications, in the field of high-performance
computing data is usually generated on specialized, expensive, and highly
sought-after hardware.  We therefore need to consider data produced on HPC
systems as valuable and difficult to come by.  Moreover, scientific data
generalizes poorly, e.g.  different applications will use slightly different
PDEs, as well as modeling and temporal resolutions. Hence, it is hard to
estimate in advance how big a training set needs to be in order to accelerate a
particular problem.

First we address the issue of trust by treating the trained model as an
\emph{in situ} component of the solver. Iterative solvers give us the unique
opportunity to to do this in a strait forward way: we can always use the
``final iteration'' of our solver to ensure that the PDE is satisfied. With
regards to the expense of the collected data, the approach we propose also
deals with this challenge. We have implemented a real-time ``online'' learning
algorithm where all relevant data is localized to the problem being solved. Our
code keeps a history of the time steps being simulated and gradually grows the
training set. Consequently, there is no need to pre-compute massive data sets
that sufficiently span a wide range of the solution space so that the learned
model would (hopefully) generalize well.

In a nutshell, by using deep learning as a flexible \emph{in situ} accelerator
for iterative methods in real-time rather than as a means to fit a surrogate
model, we are able to harness the intrinsic strength of deep learning and
dedicated hardware/software to produce useful science without the typical
challenges common to machine learning in computational science.

\subsection{A Class of Problems and Their Computational Cost}

In a PDE-based simulation, one usually starts with a time-dependent partial
differential equation (or a system of such equations) on some domain $\Omega$,
with some initial and boundary conditions. A simple example of such a problem
is
\begin{align}
    \frac{\partial u\left(\mathbf{x},t\right)}{\partial t}=R(u\left(\mathbf{x},t\right)),\\
    u(\mathbf{x},0)=g(\mathbf{x}) ,\ \mathbf{x}\in \Omega, \\
    u(\mathbf{x},t)= h(\mathbf{x},t)+\frac{\partial u(\mathbf{x},t)}{\partial n} ,\ \mathbf{x}\in \partial\Omega, 
\end{align}
where $R(u)$ is some operator (which can be linear or nonlinear), $g$ is some
initial condition, and $h$ is the prescribed value on the boundary of the
solution along with the normal derivative of the solution along the boundary of
the domain.  To numerically solve this problem, one discretizes in space and
time.  The choice of these discretizations depend on factors such as desired
accuracy of the solution, scales of interest in the problem, and mathematical
properties of the underlying solutions among many factors. Regardless of the
method chosen, the problem at hand, and the desired accuracy, at some point a
solution to a linear system of equations which takes the form:
\begin{align}
    \mathbf{A}x=\mathbf{b},
\end{align}
is required. In practice, as both the desired resolution $n$ and dimension $d$
of the problem grows, the size of this linear system is $O(n^{2d})$ and becomes
so large that it is infeasible to explicitly store the matrix $A$, let alone
use exact direct solution methods that require $O(n^{6d})$ operations to attain
a solution~\cite{NATrefethen}. As this is infeasible for any realistic $n$,
iterative methods for linear problems are standard because they require at most
$O(n^{4d})$ number of operations to reach machine precision and can be stopped
far earlier to a desired accuracy. A successful class of these iterative
methods are known as Krylov Subspace methods \cite{NATrefethen}. In general,
these methods work by projecting the original  $O(n^{2d})$ dimensional problem
into a lower dimensional Krylov subspace $K_m(\mathbf{A},\mathbf{b}) \subset
\mathbb{R}^{n^d}$  spanned by the Krlov sequence
$\{\mathbf{b},\mathbf{A}\mathbf{b},\mathbf{A}^{2}\mathbf{b},\dots,\mathbf{A}^{m-1}\mathbf{b}\}$.
The convenience and efficiency of these methods is particularly clear when one
takes notice of the fact that the Krylov sequence can be formed by repeatedly
applying the matrix operator rather than explicitly needing to construct and
store the matrix $\mathbf{A}$ itself
$\{\mathbf{b},\mathbf{A}(\mathbf{b}),\mathbf{A}(\mathbf{A}(\mathbf{b})),\mathbf{A}(\mathbf{A}(\mathbf{A}(\mathbf{b}))),\dots\}$.
While many successful methods for linear systems fall under Krylov subspace
methods such as Generalized Minimal Residuals (GMRES), Conjugate Gradient (CG),
Biconjugate Gradients (BCG) and its many variants \cite{NATrefethen},  we focus
on GMRES to illustrate our data-driven approach to accelerate a simulation.  In
particular, we focus on a particular form of GMRES known as $k$-step restarted
GMRES or GMRES($k$) that restarts GMRES iterations using the final previous
GMRES iteration as new initial guess. Restarting the algorithm is essential
when the number of operations and storage requirements are prohibitively large.
In this work, we focus on accelerating GMRES($k$) as it is a common choice to
solve the Poisson equation~--~or in the case of 3-dimensional problems, the
Laplace equation \cite{cai2014}.

While GMRES is a powerful method, like any iterative method, a good initial
guess can vastly improve performance of the method.  That said, it is well
known that initial guesses that do not incorporate problem related information
can negatively impact the performance of GMRES \cite{KrylovMethodsAnalysis}.
However, with deep learning we aim to show that we can learn effective initial
guesses that accelerate GMRES in real-time, effectively acting as left
preconditioners since no extra work is needed once a solution is obtained.  The
reason that these learned initial guesses we supply to GMRES can effectively
act like left preconditioners lies in the fact that our model learns initial
guesses that aim to optimize both the initial residual and the rate of
convergence of GMRES simultaneously.

\subsection{Accelerating GMRES}

Historically, accelerating GMRES through good initial guesses has been
difficult to achieve and typically poses a high risk since performance can
degrade with guesses that do not somehow incorporate problem specific
information \cite{KrylovMethodsAnalysis}. Furthermore, it is standard knowledge
among practitioners that simply reducing the normalized initial error through a
naively constructed initial guess (such as a solution interpolated from a
coarse grid) will not improve convergence behavior of GMRES. However, we find
that deep learning can be used to train a model that will take in a problem and
propose an optimal\footnote{Here we use ``optimal'' to mean the highest
possible rate of convergence.} initial guess. To point out how the rate of
GMRES can be accelerated with only an initial guess, consider the Krylov space
that GMRES constructs when a non-zero initial guess $x^0$ is used.  In this
case, the initial residual is given by $r^0=\mathbf{A}x^0-b$ rather than just
$b$ as demonstrated earlier. Then, the $n$-th Krylov subspace is given by
$K_n(\mathbf{A},r^0)=\text{span}\{r^0,\mathbf{A}r^0,\mathbf{A}^{2}r^0,\dots,\mathbf{A}^{n-1}r^0\}$.
Since the structure of the initial residual has the potential to influence
every Krylov space element, a good initial guess that encodes problem related
information has the potential to not just reduce the initial residual
$r^0=\mathbf{A}x^0-b$, but also simultaneously accelerate the rate at which
GMRES converges.

We emphasize here that initial guesses have the potential to improve
\textit{both} the initial residual $r^0$ and the rate of convergence of GMRES.
Improving the initial residual $r^0$ using a neural network is straightforward,
however improving the rate of convergence of GMRES using deep learning is more
subtle. The primary reason for this is that a guess that improves the initial
residual $r^0$ does not necessarily imply that GMRES will solve the problem at
an accelerated rate.  In the approach we outline here, we construct a neural
network that encodes the problem, and use the fact that we are working in
real-time to be able to asses whether initial guesses provided by the neural
network are improving the rate of convergence. To be more precise, we monitor
the performance in order to be able to use real-time performance metrics to
decide what data should be added to the training set and when the neural
network should be trained in order to accelerate performance.

\subsection{An AI-Accelerated Poisson Solver}
\label{sec:poisson-solver}

The Poisson equation 
\begin{equation}
\nabla^2 u(x, t) = f(x, t),
\label{eq:poisson}
\end{equation}
covers a vast set of applications ranging from electromagnetism, heat
diffusion, quantum mechanics, and hydrodynamics. Different applications are
encoded in the source term $f$, the boundary conditions, and the dimensions of
$x$ and $u$.

\begin{figure}
    \centering
    \includegraphics[width=\textwidth]{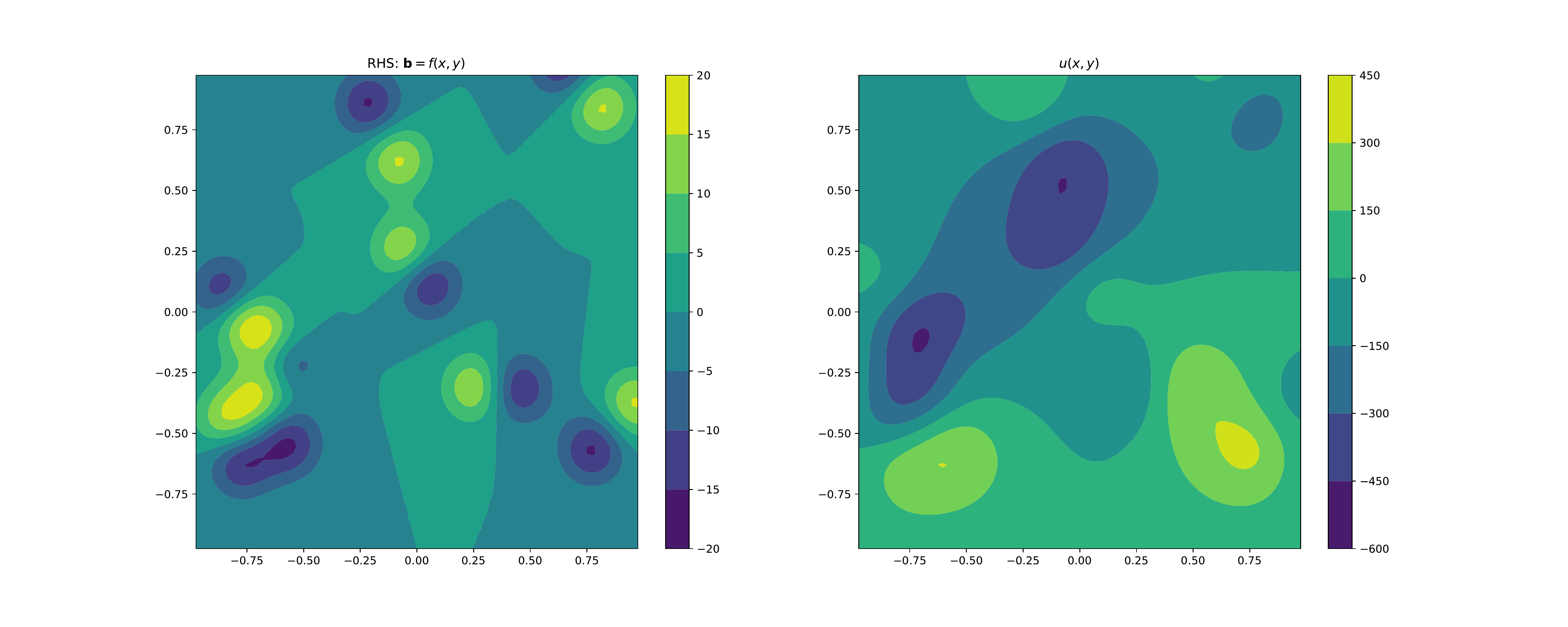}
    \caption{Example of the Poisson problems studied here -- corresponding to
    positively and negatively charged particles randomly placed in the
    simulation domain. The left panel shows a random charge distribution. The right
    panel shows the corresponding solution to Eqs.~(\ref{eq:poisson}) and
    (\ref{Poisson}).}
    \label{fig:poisson}
\end{figure}

As a starting point, we will assume that $f$ is localized (has finite support),
that $u, t\in\mathbb{R}$ and that $x\in \mathbb{R}^2$. Discretizing
Eq.~\eqref{eq:poisson} yields a linear problem of the form shown in
Eq.~\eqref{Poisson} (where $\mathbf{A}$ is a discretization of $\nabla^2$ and
$\mathbf{f}$ is a discretization of $f(x,t)$). In this form, the Poisson
equation can be used to describe electromagnetic interactions between charged
particles. An example of such a Poisson problem corresponding to randomly
distributed charged particles is shown in Fig.~\ref{fig:poisson}. In this case,
the solution to Eq.~\eqref{eq:poisson} corresponds to the electric potential.

In the next section we demonstrate that by going to three dimensions,
Eq.~\eqref{eq:poisson} can be used to describe hydrodynamic flows.
Consequently, accelerating an iterative linear solver as discussed earlier
holds great potential for reducing the computational cost of the problems we
are targeting.  To this end, since Poisson-type problems form the basis of the
problems we hope to solve, we establish a foundation by developing an
AI-powered GMRES accelerator for the Poisson problem. In particular we focus on
the Poisson problem in two-dimensions on a square domain of the form
\begin{align}\label{Poisson}
    \mathbf{A}u=\mathbf{f},
\end{align}
where we now specify that $A$ is the discrete Laplacian with corresponding
boundary conditions, $\mathbf{f}$ corresponds to the physical forces in the
problem, and $u$ is the corresponding solution.  Since we plan to use deep
learning to produce initial guesses $x_0$ to accelerate GMRES, we take a moment
to examine Eq.~\eqref{Poisson} and consider how a neural network can be
structured to learn a solution. Since we always have the RHS of
Eq.~\eqref{Poisson} available (even if we do not yet know $u$), we structure
the network such that the RHS of the Poisson problem is the neural network
input and the solution is the network output. So, we seek to construct a neural
network of the form 
\begin{align}
    u= N(\mathbf{f}).
\end{align}
Since we plan to use this neural network to accelerate an ongoing simulation in
real-time, efficiency with computing resources and memories are of fundamental
importance: we plan to construct a neural network $N$ that is fast to train
and economical with memory.

We can use the formal solution of Eq.~\eqref{eq:poisson} to make an informed
guess about which properties of the neural network are important. If the formal
solution can be represented by the network, we conjecture that there is a good
chance that a network with few parameters can be trained quickly. A formal
solution on some domain $\Omega$ can be expressed as
\begin{align}
    u=\int\limits_{\Omega }{g(\mathbf{x},\mathbf{y})f(\mathbf{y}), d\mathbf{y}} ,
\end{align}
where $g$ is the Green's function.  We note that for general boundary value
problems the solution does not necessarily depend on the source term $f$
through a convolution if the problem is not translationally invariant since in
this case the Green's function $g$ itself may not be translationally invariant.
Consequently, we should think of the solution as a non-local problem since we
technically have a Fredholm integral equation of the first kind relating the
solution to the RHS. Likewise, if we consider the discrete problem then we see
that the solution depends on the RHS through  $u=\mathbf{A}^{-1}\mathbf{f}$.
As is well known, the matrix inverse of $\mathbf{A}$ is dense even if the
discrete Laplacian is itself sparse. Consequently the key thing to notice is
that the relationship between $\mathbf{f}$ and the solution $u$ is inherently
non-local, so we must take special care of this when designing an appropriate
neural network.

\subsection{Relation to Fluid-Flow Problems}

To illustrate how equations of the form of Eq.~\eqref{eq:poisson} are related
to a wider class of problem, we show that simply by going to 3-dimensions, we
describe a whole new class of physics: fluid flow. We introduce the vector
velocity field $\mathbf{u}(x,y,z)$, the applied body forces
$\mathbf{f}(x,y,z)$, and the scalar pressure field $p(x,y,z)$ (all defined on
some three-dimensional domain $\Omega$).  The equations for Stokes flow take
the form 
\begin{align}\label{StokesEqn}
    \mu\nabla^2 \mathbf{u}-\nabla p + \mathbf{f}=0 \\
    \nabla \cdot \mathbf{u} =0, \nonumber
\end{align}
where $\mu$ is the dynamic viscosity. In a number of ways, our target problem
is intimately tied to the standard Poisson problem both in a mathematical sense
and in the sense that numerical methods for the Poisson equations overlap with
the methods for the equations for Stokes Flow.  First, writing the
non-divergence condition equation of Stokes flow component-wise, we have 
\begin{align}
    &  \nabla^2 {u}_i = \frac{1}{\mu}\left( \frac{\partial p}{\partial {{x}_{i}}}-{{{f}}_{i}} \right) \, , \, i=1,2,3.
\end{align}
Consequently, we can think of the Stokes equation in three-dimensions as
essentially being three coupled scalar Poisson equations where the solutions
satisfy the divergence free condition. In practice the velocity for Stokes flow
(and for many fluid flow problems in general) is often related to pressure
through essentially yet another Poisson equation.  While  the subtleties of how
one relates the pressure and velocity in practice depends highly on the
underlying numerical method being used (i.e projection methods, pressure
correction methods, etc), in principle one arrives at a variation of the
following manipulations of the governing equations in question. We illustrate
this for the Stokes flow Eq.~\eqref{StokesEqn}: by taking the divergence of the stokes flow
equation in Eq.~\eqref{StokesEqn}, making use of the incompressibility
condition, and formally interchanging the divergence and Laplacian operators, we
arrive at the following scalar Poisson equation for pressure
\begin{align}
    \nabla^2 p = \nabla \cdot \mathbf{f}.
\end{align}


\section{A Real-Time AI-Based Acceleration Algorithm}

A time-dependent simulation is essentially a sequence of linear problems of the
form $Ax_i=b_i$ being solved over the course of a simulation, e.g. once per
time step, giving us a ``time series'' of length $T$. We refer to this sequence
as our ``simulation''.  The goal here is to train a neural network $N(b)$ in
real-time as linear problems are solved by GMRES($k$). This is accomplished by
having the neural network provide an initial guess $N(b_i)=x_i^{0}$ to
GMRES($k$).  The training objective is that this initial guess then improves
the rate of convergence of GMRES($k$).

The data set used to train the model consists of RHS-solution pairs $\{
\left(b_i, x_i \right)\}$. However, unlike traditional deep learning
approaches, the goal here is to train the network in real-time while data is
being generated from the simulation. This naturally leads to an online
supervised learning problem since at a given time $t<T$, we only have access to
$N(t)$ number of $\left(b_i, x_i \right)$ pairs. In the formulation presented
here, at time $t<T$, the training set is $X_t=\{ \left(b_i, x_i
\right)\}_{1}^{N(t)}$.  In order to ensure a high-quality data set some time
steps are discarded, while only ``high-quality'' ones are kept (this is
crucial, and is discussed in a later section).

\begin{figure}
    \centering
    \includegraphics[width=0.49\linewidth]{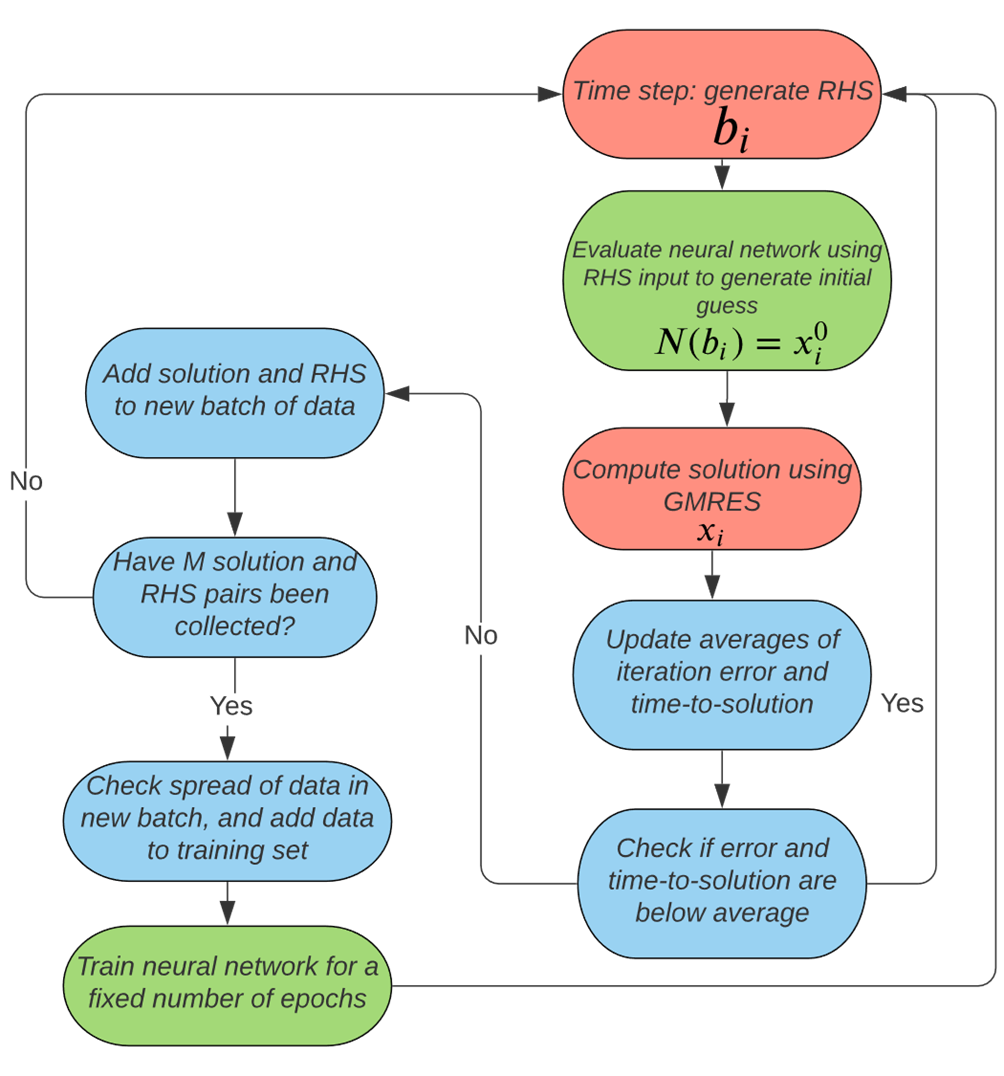}
    \caption{
        Experiment for accelerating GMRES with neural networks in real-time.
        Red boxes indicate traditional computational steps, blue boxes data
        collection, and green boxes indicate deep learning.
    }
    \label{fig:chart}
\end{figure}

We train the neural network for a fixed number of epochs during the training
phase using batched gradient descent where samples are randomly selected from
$X_t$. When $A$ is the discrete Laplace operator, networks that are able to
learn non-local relationships work especially well.  Good performance was
obtained with 4-6 layer CNNs with sufficiently large and dilated kernels, as
well as the more sophisticated FluidNet~\cite{FLuidNet} architecture described later.

Quality metrics such as the 2-norm of the GMRES residual at a particular
iteration and rate-of-convergence for each guess provided by the neural network
are tracked as the simulation runs.  After $M$ solutions have been computed
these metrics are used to decide which of the $\{ \left(b_i, x_i
\right)\}_{1}^{N(t)}$ pairs are added to the training set for the next training
phase (at some time $t'>t$).
The flow chart provided in Figure \ref{fig:chart} outlines our real-time
learning algorithm.

\subsection{Neural Network Architecture}\label{Secion_NetArch}

We discussed the properties necessary to capture solutions of the Poisson
equation in Section~\ref{sec:poisson-solver}.  Convolutional neural networks
(CNN) satisfy these and are well known to be a useful tool. However, there are
two key issues that require particularly careful treatment when selecting an
appropriate neural network architecture for our real-time learning context.
\begin{enumerate}
    \item  The neural network must be capable of learning non-local
        relationships without compressing resolution of data
    \item  The network should be relatively ``easy'' to train in the sense that
        the training of the network should not be on a time-scale comparable to
        that of the simulation
\end{enumerate}
The first key issue stems from the fact that we are primarily interested in
boundary value problems on finite domains that are not periodic. Even in the
case of a linear problem, the relationship between the solution and the right
hand side is given by an integral over the entire domain
$\Omega$. Consequently, it is not surprising that in general we can expect
our network to perform best if it can capture non-local information. Unfortunately, this
necessity is at odds with the fact that convolutional neural network layers
usually use kernels that are inherently localized. In this
case, the receptive field of these networks can only grow with the depth of the
neural network. While relying on an increasing network depth may not sound too
taxing, it is a less than ideal solution. Deeper networks are slower to train,
and the necessary depth to compensate for higher resolution problems would lead
to a fairly poor trade-off in performance. Personal experience has also shown
that allowing convolutions to shrink spatial resolution so that non-local
features are learned, and then up-scaling to the desired output
resolution yields poor results. This likely happens because important
information that GMRES needs is lost as the convolutions reduce the spatial
resolution.

\begin{figure}
    \centering
    \includegraphics[width=0.95\linewidth]{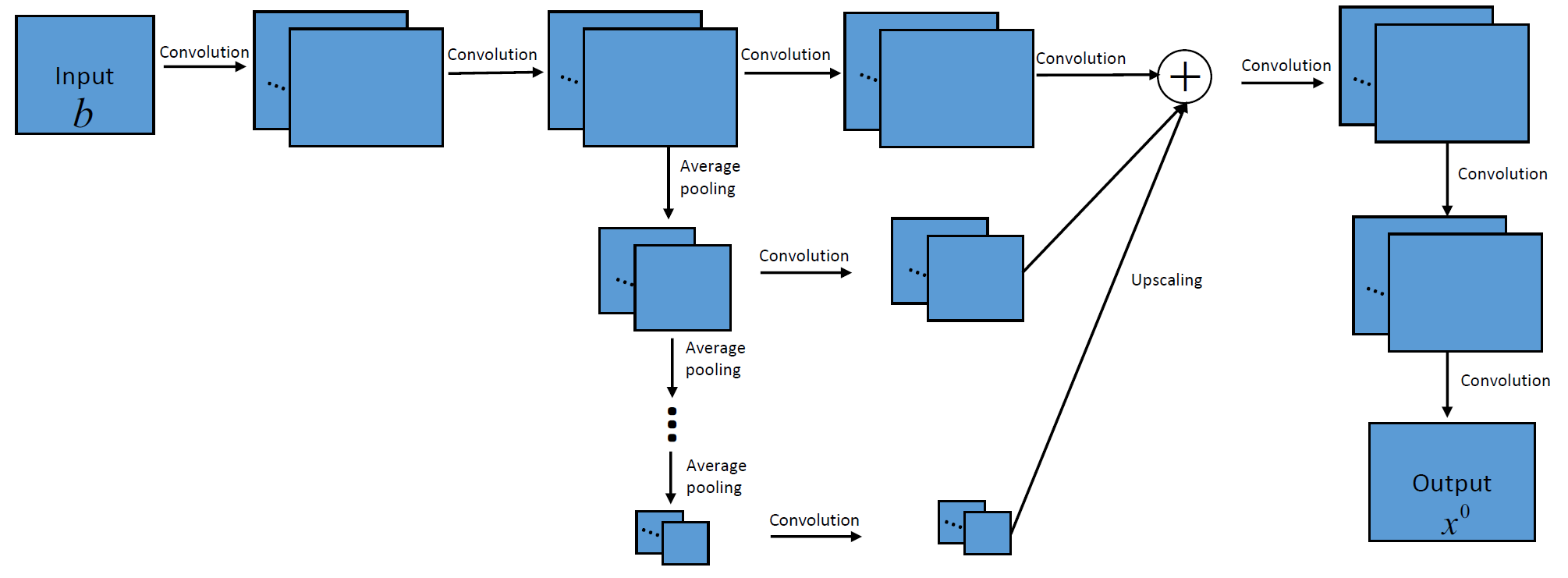}
    \caption{
      We use a variation of the FluidNet pressure model of \cite{FLuidNet} in
      our neural network architecture shown in Figure \ref{fig:Resblocks}. In
      the model used here, the number of average pooling operations depends on
      the problem at hand rather than being fixed.
    }
    \label{fig:Fluidnet}
\end{figure}

For our test problems, we have found a particular network
architecture that stands out in its ability to encode the desired non-local
information while still being an all convolution architecture that does not
need to rely on network depth to encode non-local information.  This network is
the FluidNet architecture's pressure component from
\cite{FLuidNet}~--~shown in Fig.~\ref{fig:Fluidnet}. This network
essentially works by down-sampling the input
using $2^d$ average pooling a number of times while retaining the original and
each down-sampled field.  Each of these fields with different resolutions are
then passed through a series  of convolutional layers in parallel, and then all
are up-sampled to the original resolution and summed.  This sum is then passed
through a series of convolutions to then produce the output field.  In the
architecture we use here, we essentially use a variation of this idea and
increase the depth of this network by stacking residual blocks before the
output layer.

The second key issue stems from the fact we are working in real-time:
we must not slow down the regular (unaccelerated) code with long training times.
Consequently, we are far more interested in a model that converges to a useful
output quickly (and not necessarily highly accurately) than a model that slowly
converges to a highly accurate output eventually. 
While shallower networks
are known to converge to useful results quickly, their accuracy is inferior to
deeper networks. Conversely, deeper networks converge slowly compared to
shallow networks but are more accurate.

\begin{figure}
    \centering
    \includegraphics[width=0.85\linewidth]{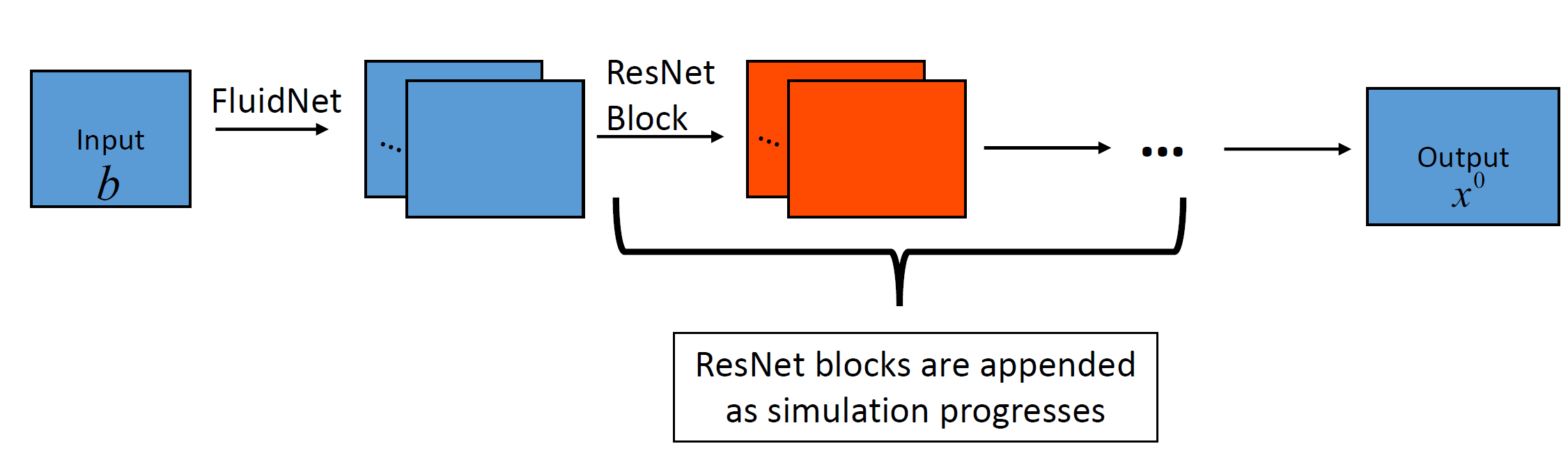}
    \caption{
       Resnet blocks are appended to the network before the output layer as the
       simulation runs. By starting with a shallow model and building up to a
       deeper model, we are able to benefit from fast convergence of a shallow
       model during early (low data) stages of a simulation and the
       higher accuracy of deeper models during later stages (more data) of a
       simulation.
    }
    \label{fig:Resblocks}
\end{figure}

To balance network training speed with network depth, we have taken the
approach of adaptively adding residual blocks to the output of
FluidNet as the performance of the
network begins to stagnate.  A diagram of this approach is illustrated in
Figure \ref{fig:Resblocks}.  With this approach, the relatively shallow network
converges to a model that is immediately useful, and then is allowed to grow
into a model that can provide outputs of improved quality. Furthermore the
real-time nature of the problem restricts initial training sets to a small
size.  Naturally, there is a high likelihood that the state of the training set
encountered initially will not be representative of all data encountered
over the course of the simulation.  Consequently there is a high likelihood
that an initially deep network will over-fit and perform poorly as the
simulation progresses.  By starting with
an initially shallow model, we are able to mitigate these issues.

\subsection{Online Data Collection Process}

A key part of our approach relies on collecting data from the simulation
selectively. There are multiple reasons to carefully collect data selectively
in this real-time context. First, since we are not working with a
representative distribution of data initially\jpb{}{~--~}and are not
necessarily guaranteed to have a representative distribution of data at any
point in the simulation\jpb{}{~--~}we must be sure that certain types of
samples are not significantly over-represented in the data set and over-fit by
the model. Similarly, spurious data points can also negatively affect training
and model performance significantly.  Furthermore, since we cannot take
an arbitrarily long time to train the model (we must keep the training
time-scale relatively small compared to the simulation time scale), using a smaller
representative data set rather than a massive, highly redundant data set can be beneficial.

Every simulation starts with an initial small data set of $N$ solution and RHS
pairs $X_t=\{(b_i,x_i) \}_{i=0}^N$ by some time $\tau_0$ and we use a
neural network $N(b)$ trained on this training set $X_t$ to produce initial
guesses $x^0=N(b)$. For $t>\tau_0$, as the simulation progresses and before
every GMRES($k$) iteration, the initial guess is generated by a forward
evaluation of the neural network with the RHS as an input, \emph{i.e.} we have
$x^0=N(b)$. This guess is then used to start the GMRES($k$) iterations. Through
training the neural network with a loss function  that maps the RHS to
solutions and the evaluation of the initial residual $x^0=A N(b)-b$, it is easy
to determine whether this guess reduces the initial error. However, what is
more important and not as easy to directly ascertain is whether this guess
accelerates GMRES($k$). To this end, we keep track of two metrics and a moving
average of these metrics.  The first metric is the GMRES($k$) residual at the
end of the first restart (\emph{i.e.} the $k$th iteration) with the provided
initial guess:
\begin{align}
    E_k(x^0)=\mathbf{A}x^k-b .
\end{align}
The second metric is the time to reach the solution 
\begin{align}
    \text{TOS}(x^0)=\text{Wall-clock time taken for GMRES to converge to $x=\mathbf{A}^{-1}b$ within tolerance.}
\end{align}
Both $E(x^0)$ and $\text{TOS}(x^0)$ are naturally related to the rate of
convergence of GMRES. 
To
make use of these metrics, we use a moving average to set a meaningful baseline
of what constitutes ``good"  values of $E(x^0)$ and $\text{TOS}(x^0)$. We
define this unweighted moving average using a window that only looks at the
last $p$ number of steps. 
Explicitly, when we have $N$ samples of some
quantity $y$, we define this moving average as
\begin{align}
    M_p(\{y_i\}_{i=0}^{N})=\frac{y_{N-p}+\dots+y_N}{p}.
\end{align}
Algorithm~\ref{al:luna} summarizes how we collect data at some time step
$t=t_n>\tau_0$, we compute a solution $x^n$ using an initial guess $x_n^0$.
Then we ccalculate $E_k(x_n^0)$ and
$\text{TOS}(x_n^0)$.
We also compute the moving averages of these quantities as
previously defined including the most recent data
$M_p\left(\{E_k(x_i^0)\}\right)$ and $M_p\left(\{ \text{TOS}(x_i^0)\}\right)$.
We then compare the current error and time to the mean and we check
if the following two conditions hold:
\begin{align}
    E_k(x_n^0) < M_p\left(\{E_k(x_i^0)\}\right) \\
    \text{TOS}(x_n^0) < M_p\left(\{ \text{TOS}(x_i^0)\}\right).
\end{align}
In particular we check if the current guess provided by the network achieved a
better than average error by the $k$th iteration and a better than average
wall-clock time to solution. If the initial guess did meet both criteria, we
simply proceed to the next time step and repeat what has been described so far.
However, if the initial guess did not meet both criteria, then the model
under-performed on this particular data.  Consequently, the corresponding
solution and RHS pair $(b_n,x_n)$ is then added to a set $C_t$ containing
the candidate data points that may be added to the training set $X_t$. However
before adding the data to $C_t$, the spread between $x_n$ and all elements of
$C_t$ is checked. We denote the measure of spread between solutions as
$d(\cdot,\cdot)$ and note the definitions depends on what is appropriate for the
problem. Here, we check for sufficient orthogonality between the
solution data. In particular, we check that $x_n$ is ``sufficiently" orthogonal
to all members of $C_t$. If this is the case, then the data point is added to
$C_t$.  Depending on whether we have encountered $f_r$ (the retraining
frequency) data points that fail to meet our conditions, we either repeat what has been described so far, or we proceed to train the model.

\begin{algorithm}
    \SetAlgoLined
    Simulation has reached time $t_{n-1}$ \;
    Neural network $N(b)$ has just been trained using training set $X_t$\;
    Set of candidate data pairs $C_t$ is empty, $N=0$\;
     \While{$N<f_r$}{
      $t_n=t_{n-1}+\Delta t_{n-1}$\;
      Compute RHS $b_n$\;
      Use Neural network $x_n^0=N(b_n)$\;
      Compute solution $x_n=\text{GMRES}(x_n^0)$\;
      Compute performance  metrics  $E_k(x_n^0)$ and $\text{TOS}(x_n^0)$ \;
      Compute Averages $M_p\left(\{E_k(x_i^0)\}\right)$ \text{and} $M_p\left(\{ \text{TOS}(x_i^0)\}\right)$\;

      \eIf{$E_k(x_i^0) < M_p\left(\{E_k(x_i^0)\}\right)$ and $\text{TOS}(x_i^0) < M_p\left(\{ \text{TOS}(x_i^0)\}\right)$}{
        Proceed with the simulation. 
        Set $n=n+1$ and next  $\Delta t_{n-1}$
       }{
       Compute $d(b_n,b)$ for all $b$ $\in$ $C_t$\;
       \If{$d(b_n,b)$ for all $b$ is acceptable }{
        Add $(b_n,x_n)$ to $C_t$\;}
       Set $N=N+1$\;
       Set $n=n+1$ and next  $\Delta t_{n-1}$;      }
     }
     \caption{Collecting data after model has been trained\label{al:luna}}
\end{algorithm}

\subsection{Online Training Algorithm}

In our online approach, we only train at the start of the simulation or when
$f_r$ ``under-performing" data points $(b_n,x_n)$ are encountered.
In the former case, we simply collect all data up to a reasonable point, say about $50$ time steps. Then all of these are
added to the training set $X_t$ and the neural network is trained using the
entire training set. Following this, the simulation continues while the metrics
discussed above are used to judge when we add data to $C_t$. We then collect data as
described above until $f_r$ new data points have been found and the model is
re-trained.

In a traditional online machine learning approach, one would update the model
using solely the new data from $C_t$. However, this typically comes at the cost of
interfering with previously learned information in neural networks
\cite{online}.  To mitigate this so called ``\textit{catastrophic forgetting}''
we simultaneously train a model with data from $C_t$ and sampled data from the
training set $X_t$. This way, the neural network model is not allowed to forget
all previous information since it is being actively reminded of previous
relationships through exposure to prior data from $X_t$. When training the
neural network model, we train the network to minimize the objective
\begin{align}
    L(\tilde{x},x)= ||\tilde{x}-x||_{2}^{2},
\end{align}
where $\tilde{x}$ is the initial guess provided by the network and $x^n$ is the
solution computed by GMRES($k$). The model parameters are updated using back
propagation to calculate all partial derivatives and an adaptive gradient
descent method such as ADAM and Adagrad algorithms. Since the training set
$X_t$ can become quite large, we fix the number of batch gradient descent steps
so that the training time does not grow rapidly relative to the simulation time and
randomly sample our training set while guaranteeing that recently collected
data is used to train the model. It's key to clarify here that minimizing the
objective function is half the story, and any objective function that
implicitly minimizes the initial residual of GMRES($k$) could work in
principle.  While minimizing the objective function coincides with our goal of
minimizing the initial residual, the acceleration of GMRES($k$) is obtained
through the constant monitoring of the effect of the neural network's real-time performance in accelerating GMRES($k$). Those
metrics essentially determine what data is used, and what the model should
focus on learning.


\section{Benchmarking an Accelerated Poisson Solver}
\label{sec:prototype}

In this section we demonstrate the viability of the concepts discussed
previously. We implemented an AI-accelerated GMRES($k$)~--~which we call
MLGMRES~--~by accelerating a Python implementation of the GMRES($k$)
algorithm\footnote{Code is available here:
\texttt{https://github.com/ML4FnP/GMRES-Learning}}. While this implementation
is obviously slower than a compiled language, the relative speedups over the
original algorithm should not be affected. We work in the Python language as it
allows rapid prototyping. We emphasize that none of our algorithms depend on this particular
choice in languages~--~e.g. there has been some success applying these concepts to
C++-based codes~\cite{AMReX_JOSS,cai2014}. However, this will not be covered
here.

As a benchmark that represents realistic scientific PDE problems we solve a
series of Poisson problems of the form $Ax_i=b_i$ where $A$ is the discrete
Laplacian with zero-Dirichlet boundary conditions and each $i$ is a particular
time-step.  The $b_i$ used here are essentially randomly moved localized
sources similar to those in Figure \ref{fig:LinearModel} on the square
$\Omega=\left[-1,1\right]\times\left[-1,1\right]$. This is analogous to a
2-dimensional system consisting of charged particles. In a real-world setting,
these particles would move according to an interaction force (e.g.
$\mathbf{f}(x,x) = -\nabla u(x,y)$ where $u(x,y)$ is a solution to the Poisson
equation). We move the particles randomly, making this a much
harder benchmark: the test inputs are far less correlated than they would be if
they were generated by physical dynamics. During our benchmarks, we find that on average 1 in 10 time steps are retained by Algorithm~\ref{al:luna}.

Our benchmarks were performed on a single Cori-GPU node at National Energy
Research Scientific Computing Center (NERSC). Each node has two 20-core Intel$^{\circledR}$
Skylake$^\mathrm{TM}$ processors and 8 NVIDIA$^\mathrm{\circledR}$ V100 GPUs.  For the results presented here, we
trained the model using PyTorch~\cite{pytorch} running on 1 (out of the 8 available) GPUs. We compare against the same benchmarks running on a 10-core Intel$^{\circledR}$ Core$^\mathrm{TM}$ i9 desktop computer with PyTorch training on a NVIDIA$^\mathrm{\circledR}$ RTX 2070. We did not see any noticeable performance difference. This gives us confidence that no specialized hardware (other than a reasonable GPU) is required to run our AI-accelerator.

\subsection{Understanding the AI Accelerator through a Simplified model}

\begin{figure}
    \centering
    \includegraphics[width=0.85 \linewidth]{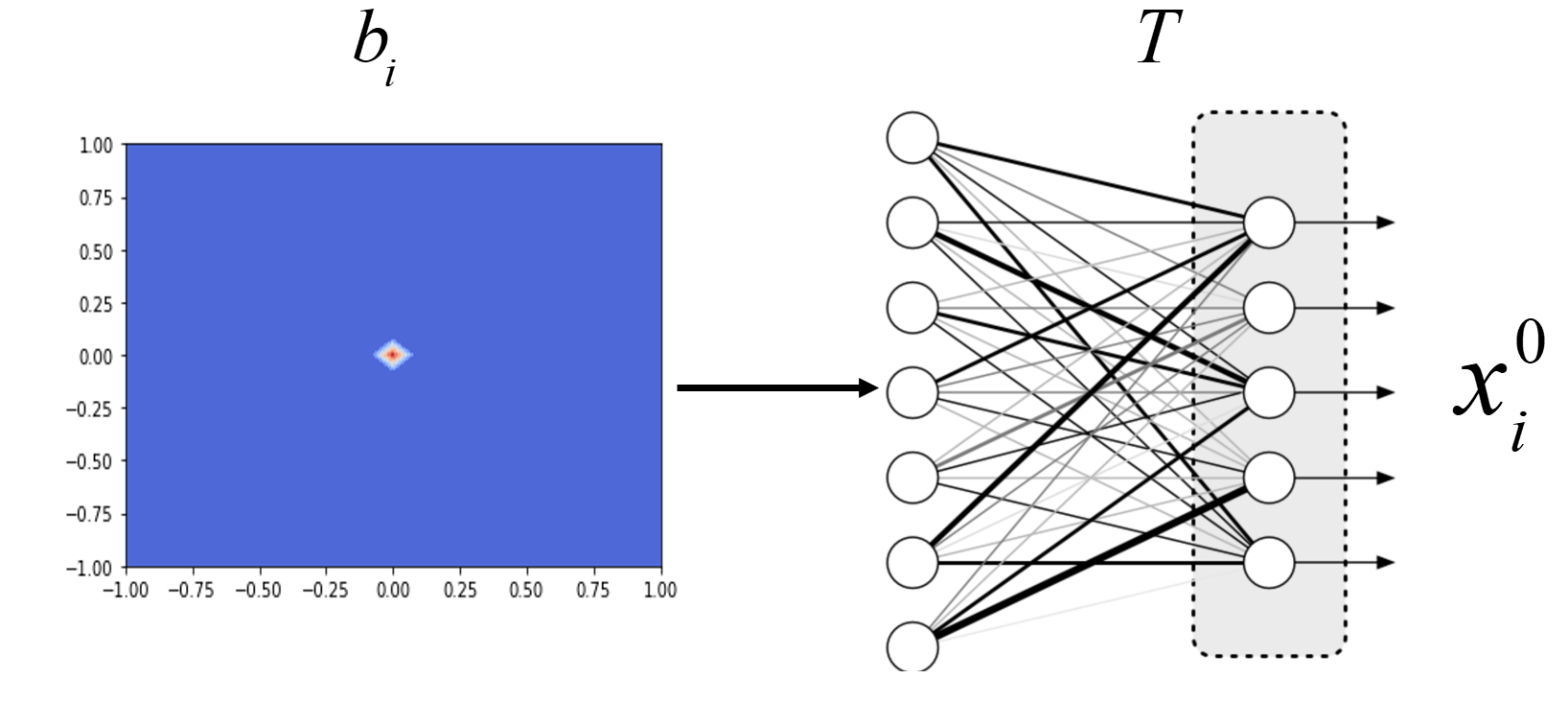}
    \caption{
      Simple single fully connected layer neural network model used to explore behavior of the accelerator.
    }
    \label{fig:LinearModel}
\end{figure}

We can gain a great deal of insight by considering a simpler neural network
architecture.  In particular if we consider a single fully connected layer as
seen in Fig. \ref{fig:LinearModel}, we can conveniently express the mapping
as a matrix operation on the input. We can express this network as matrix
multiplication of the form $Tb=x_i^0$. As $T$ is learned over the course of the
simulation, it is reasonable to expect $AT$ to have a spectrum that clusters
around the unit circle.  This is particularly clear when the norm of the
initial residual is considered. At the start of the GMRES($k$) iterations for a
particular time-step, we have $||r_0||_2=||Ax_i^0-b||_2$. However, for the
initial guess we use  $Tb=x_i^0$.  Consequently, we have
$||r_0||=||AT-I||_2||b||_2$. So we see that for the neural network to reduce
the initial residual, the spectrum of $AT$ should cluster on the unit circle.
We test this idea on a simulation where we randomly move the localized source
from Fig. \ref{fig:LinearModel} on the square with a resolution of $30 \times
30$. 

\begin{figure}
    \centering
    \includegraphics[width=0.85 \linewidth]{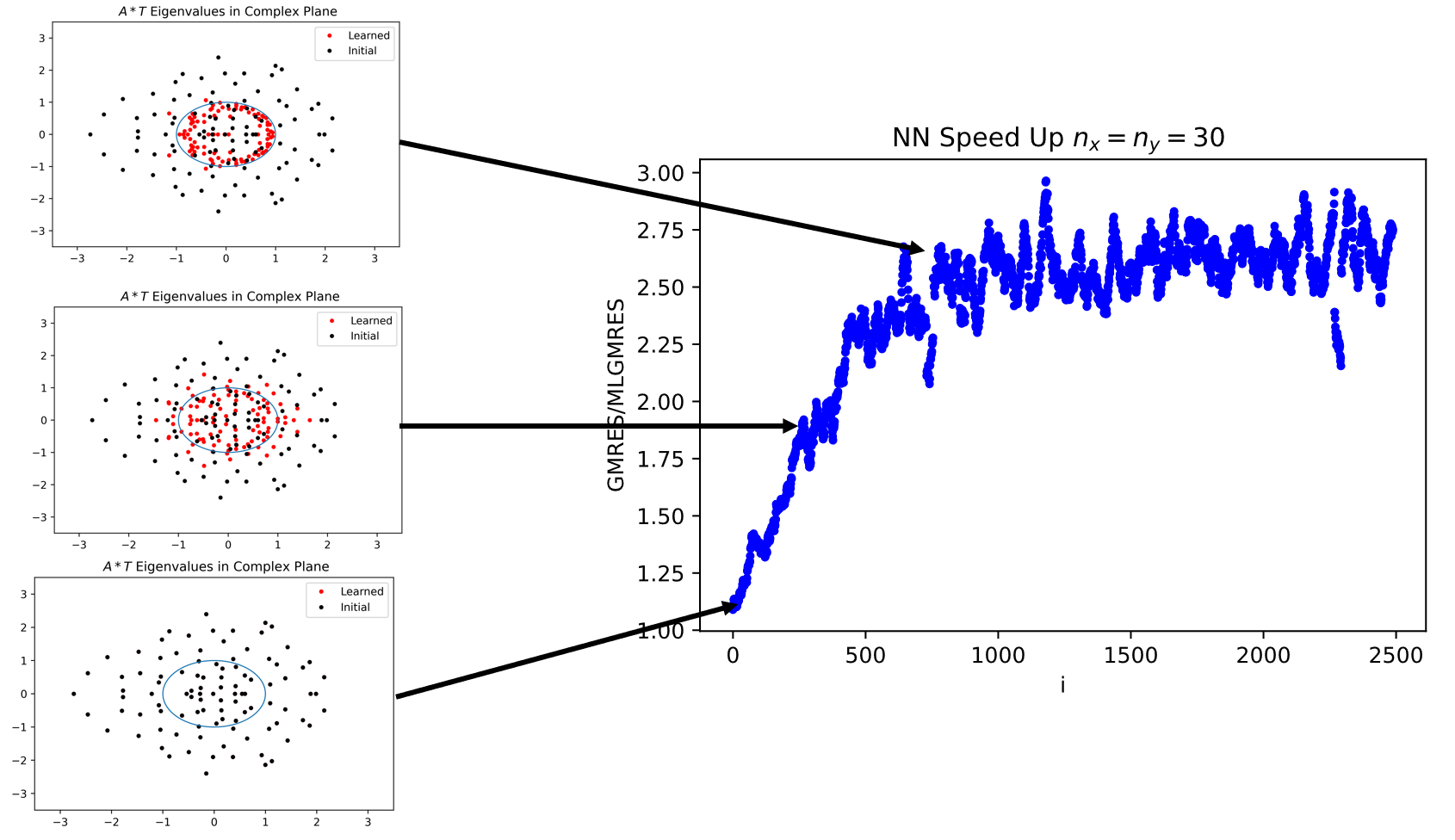}
    \caption{
      Spectrum of $AT$ over the course of a single simulation.
    }
    \label{fig:LinearTSpec}
\end{figure}

In Figure \ref{fig:LinearTSpec} we show plots of the spectrum in the
complex plane and  the speed up provided by the neural network accelerator. The
behavior observed is exactly as anticipated: we see that as we
proceed in time (steps are indexed by $i$), the neural network $T$ learns a
mapping such that the entire spectrum of $AT$ is moved onto the unit circle.
Consequently, we see that the neural network accelerator even for this simple model is able to encode information that gives $AT$ a friendlier spectrum.  This more favorable spectrum yields a lower
$||r_0||_2$ and also lends some insight into why we see not just a lower
initial residual, but also an improved rate of convergence.  In particular,
since the \textit{entire} spectrum of $AT$ is eventually shifted to the unit
circle, it is reasonable to conjecture that the network is encoding information
that overall allows a more efficient approximation of the solution within a
particular Krylov subspace.

\subsection{Purely Convolutional Network}

\begin{figure}
    \centering
    \includegraphics[width=0.49 \linewidth]{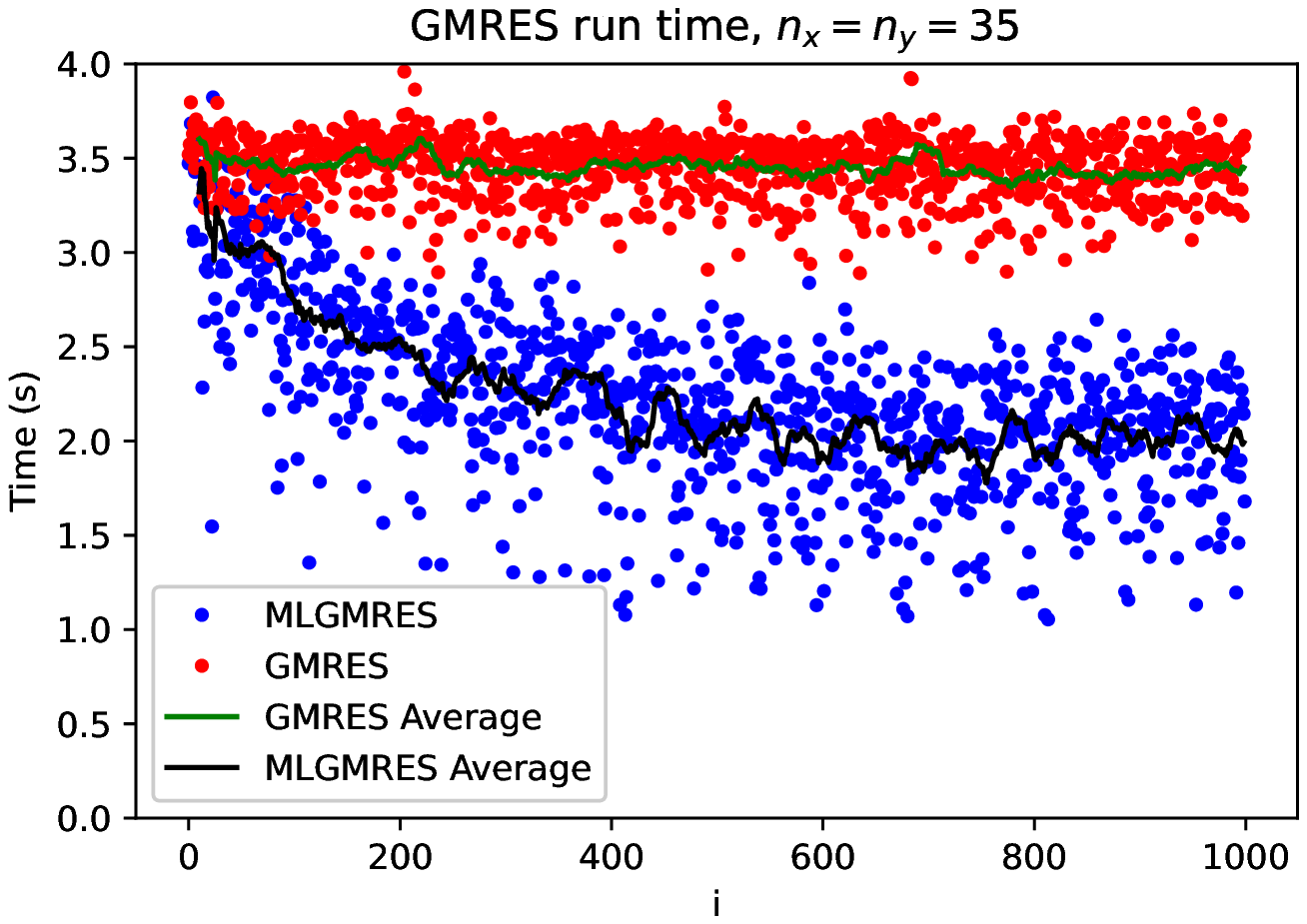}
    \includegraphics[width=0.49 \linewidth]{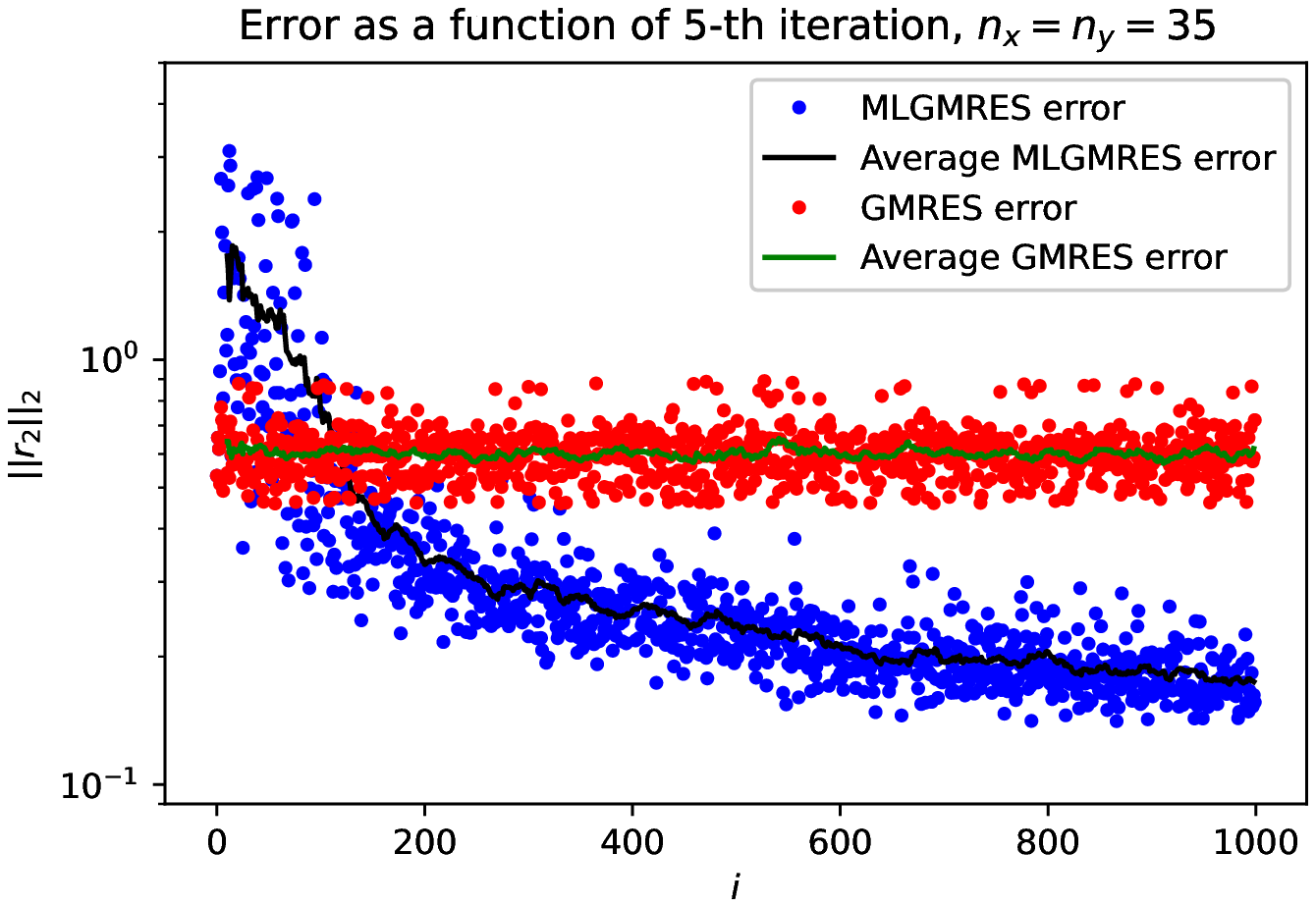}
    \caption{
        \emph{Left:} Speed up attained by MLGMRES during the simulation.
        \emph{Right:} Error of GMRES at the 5th iteration.
    }
    \label{fig:CNN1}
\end{figure}

\begin{figure}
    \centering
    \includegraphics[width=0.49 \linewidth]{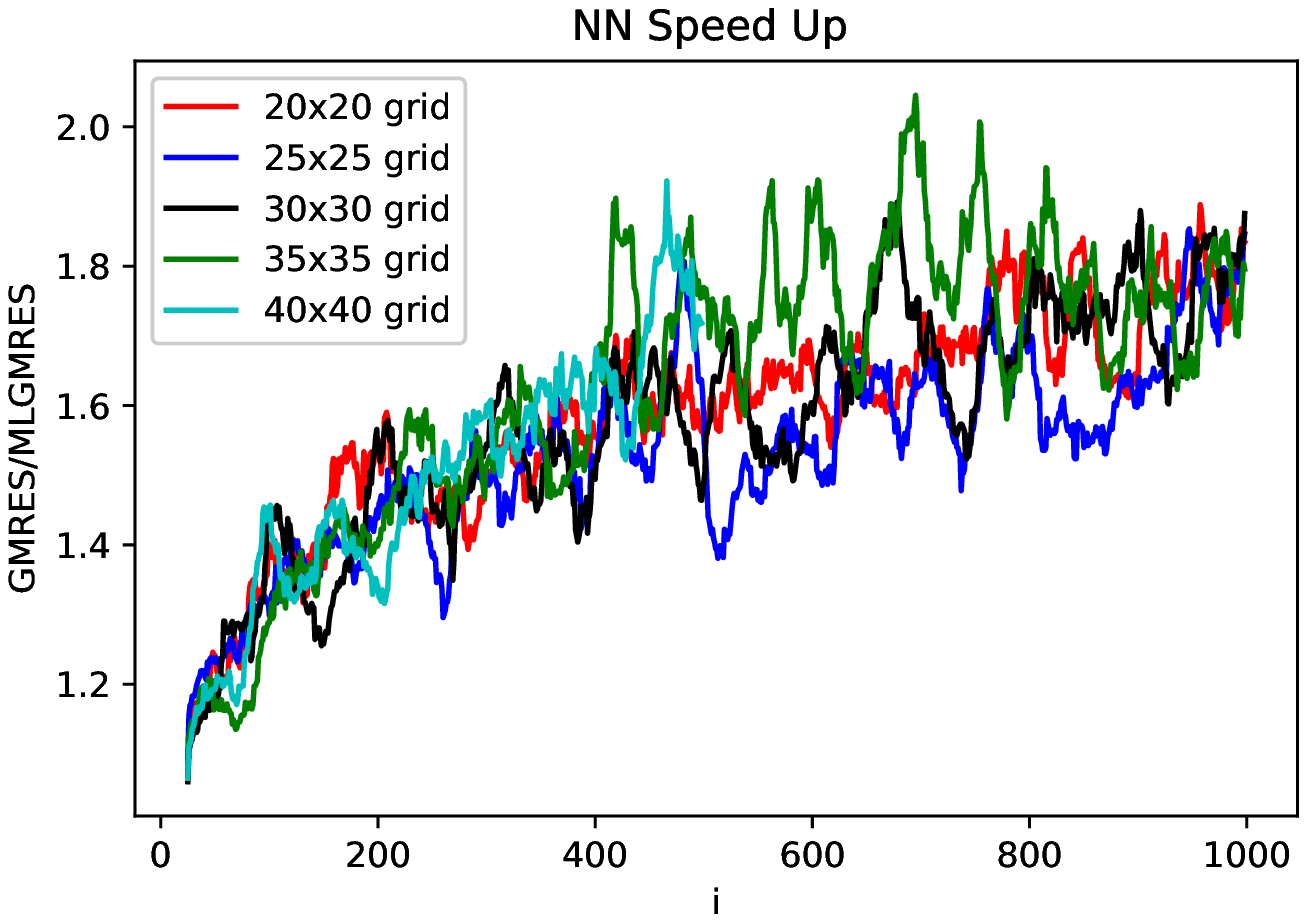}
    \includegraphics[width=0.49 \linewidth]{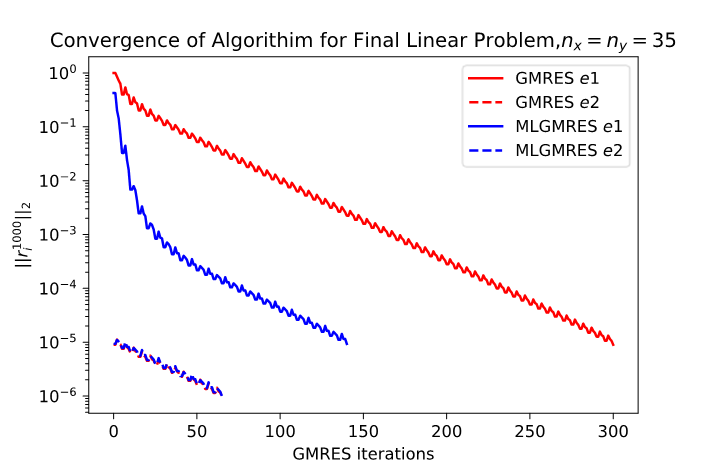}
    \caption{
        \emph{Left:} Speed up attained by MLGMRES during the simulation for different resolutions.
        \emph{Right:} Improvement in convergence rate of MLGMRES after 200 time
        steps. The final training set only contains roughly 200 data pairs.
        Dashed lines indicate restarts at a higher error tolerance $e_2$ after
        the first tolerance $e_1$ is reached. We confirm that the convergence
        rate for higher tolerances $e_2$ is not affected by the neural-network's
        initial guess.
    }
    \label{fig:CNN2}
\end{figure}

In building up to the neural network architecture discussed earlier, we
explored the use of sequential convolution networks where only convolutions are
used. Though this architecture is not particularly sophisticated, it serves as a useful starting point for understanding how powerful the idea of
deep learning based accelerators is. 
First, we show some results for a fixed resolution.  In Fig. \ref{fig:CNN1}
we plot the time-to-solution (left plot) and the fifth GMRES($k$) iteration
error (right) as of function of every time-step (indexed by $i$). The primary observation is that as the
simulation progresses, the neural network model produces initial guesses that
simultaneously improve the wall-clock time-to-solution and the fifth residual
in the GMRES($k$) iteration ($||r_5||_2$).  While an improved residual at the
fifth iteration may be useful, it does not tell use precisely whether or not
the actual rate of convergence has been improved. The speed-up we are seeing
could simply be attributed to a small initial residual with zero impact on the
rate of convergence.  To investigate this further, we look at the GMRES($k$)
residuals at each iteration for the final time step ($i=1000$). This is plotted
in Fig. \ref{fig:CNN2} (right panel).  The blue curve corresponds to the
accelerated solver.  We see that the initial residual is still fairly large, roughly half of the initial residual. Consequently, improvements in the
initial residual values seem to play a fairly small roll in accelerating the
convergence. The second feature that differentiates the accelerated MLGMRES
solver is the fact that the convergence rate is temporarily accelerated and
consequently reduces the overall time-to-solution. We therefore conclude that
the acceleration from our neural network ``wears off'' with each restart of
GMRES($k$). None the less, this servers as a compelling proof of concept for
this simple problem and network, since we see that we are able to accelerate GMRES.

Because we aim to be memory efficient, we next test how well a model with a fixed
number of parameters scales to larger problem sizes. While keeping the
number of model parameters fixed, for this simple network we dilate kernels
consistently (instead of using larger kernels) with every increase in grid
resolution. While we have discussed more sophisticated ways to deal with
different scales, here we simply dilate kernels to get a
lower bound on how robust our approach is. These results are shown in Figure
\ref{fig:CNN2}(left panel) and Table \ref{tab:freq2}. We essentially get a similar
acceleration for different resolutions, and are able to keep training times
relatively small compared to the time of the entire simulations.

\begin{table}
    \centering
    \caption{
        Comparison of GMRES solver run times for different grids using the all convolution AI-acclerator
       \label{tab:freq2}
    }
    \begin{tabular}{cccccl}
        \toprule
            & $20\times20$ & $25\times25$ & $30\times30$ & $35\times35$ & $40\times40$     \\
        \midrule
            Total GMRES Time (s)            &927   &1677 & 2718 &4254  & 3541  \\
            Total MLGMRES Time (s)          &667   &1238 &1971  &3022  &5352    \\
            Total Training Time (s)         &95.1  &86.2 &95.2  &101.5 &106.1 \\
            Training Time/Total Solver Time &0.14  &0.07 &0.05  &0.033 &0.023 \\
    \bottomrule
\end{tabular}
\end{table}

\subsection{Neural Network Designed for Multiple Scales}

\begin{figure}
    \centering
    \includegraphics[width=0.49 \linewidth]{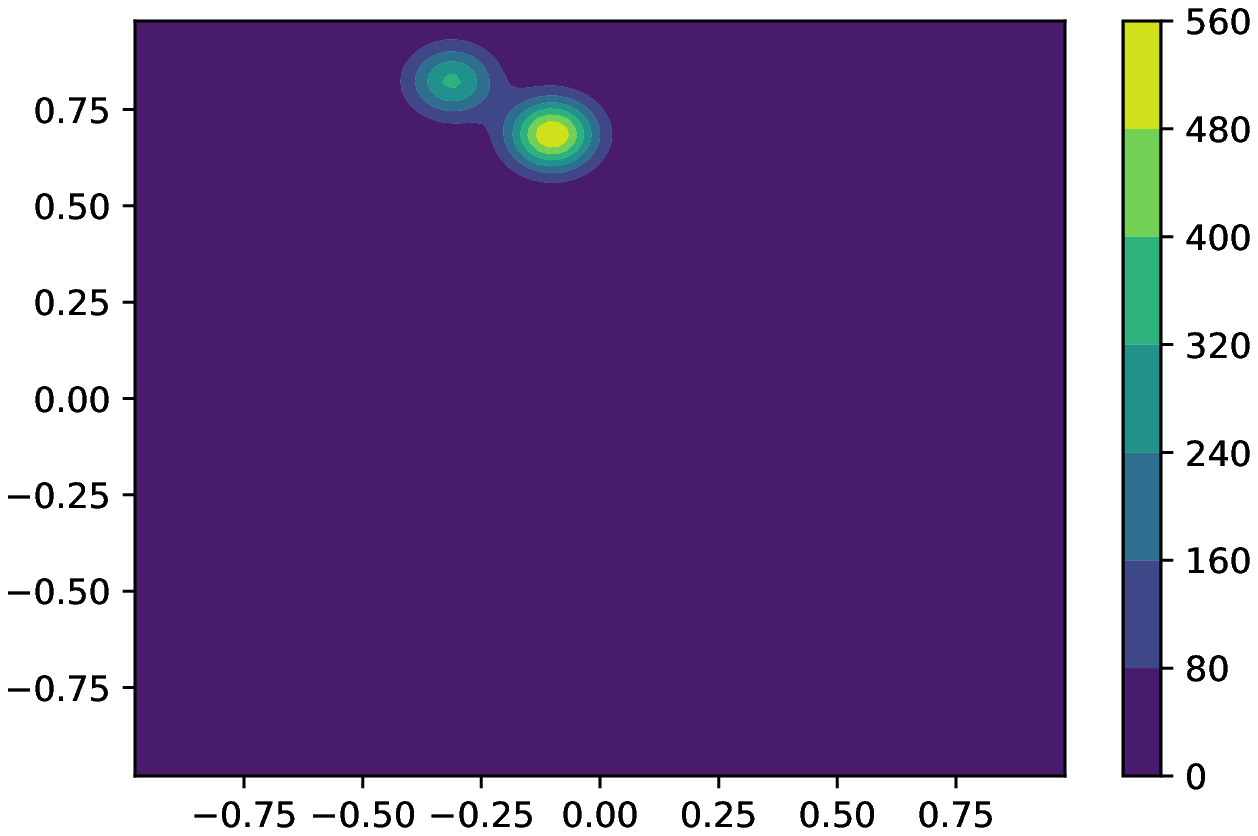}
    \includegraphics[width=0.49 \linewidth]{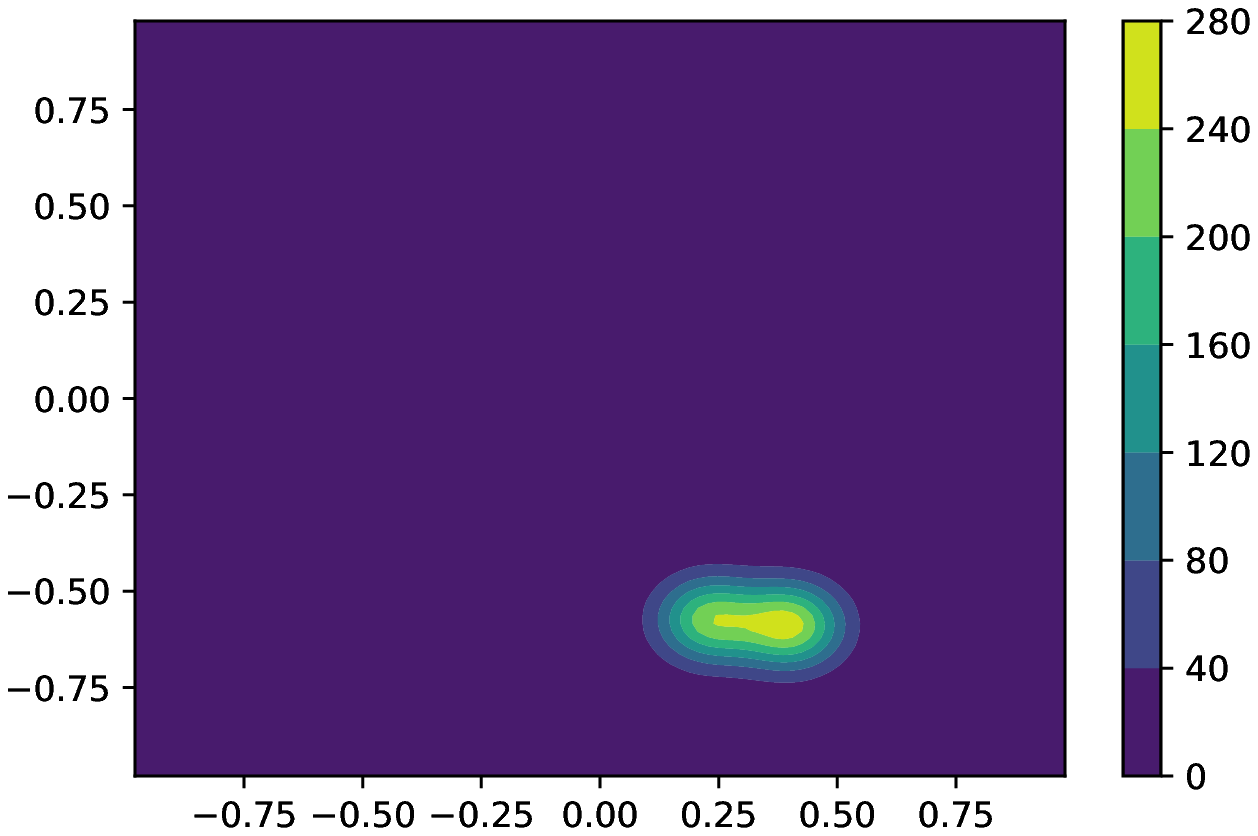}
    \caption{
        Examples of dipole forcing used for testing the MLGMRES (with adaptive FluidNet architecture) AI Accelerator. At every time step the dipole with two scales for each pole are randomly shifted (on the domain and relative to each-other) with different random overall scales at every time step.
    }
    \label{fig:DipolesExample}
\end{figure}

\begin{figure}
    \centering
    \includegraphics[width=0.49 \linewidth]{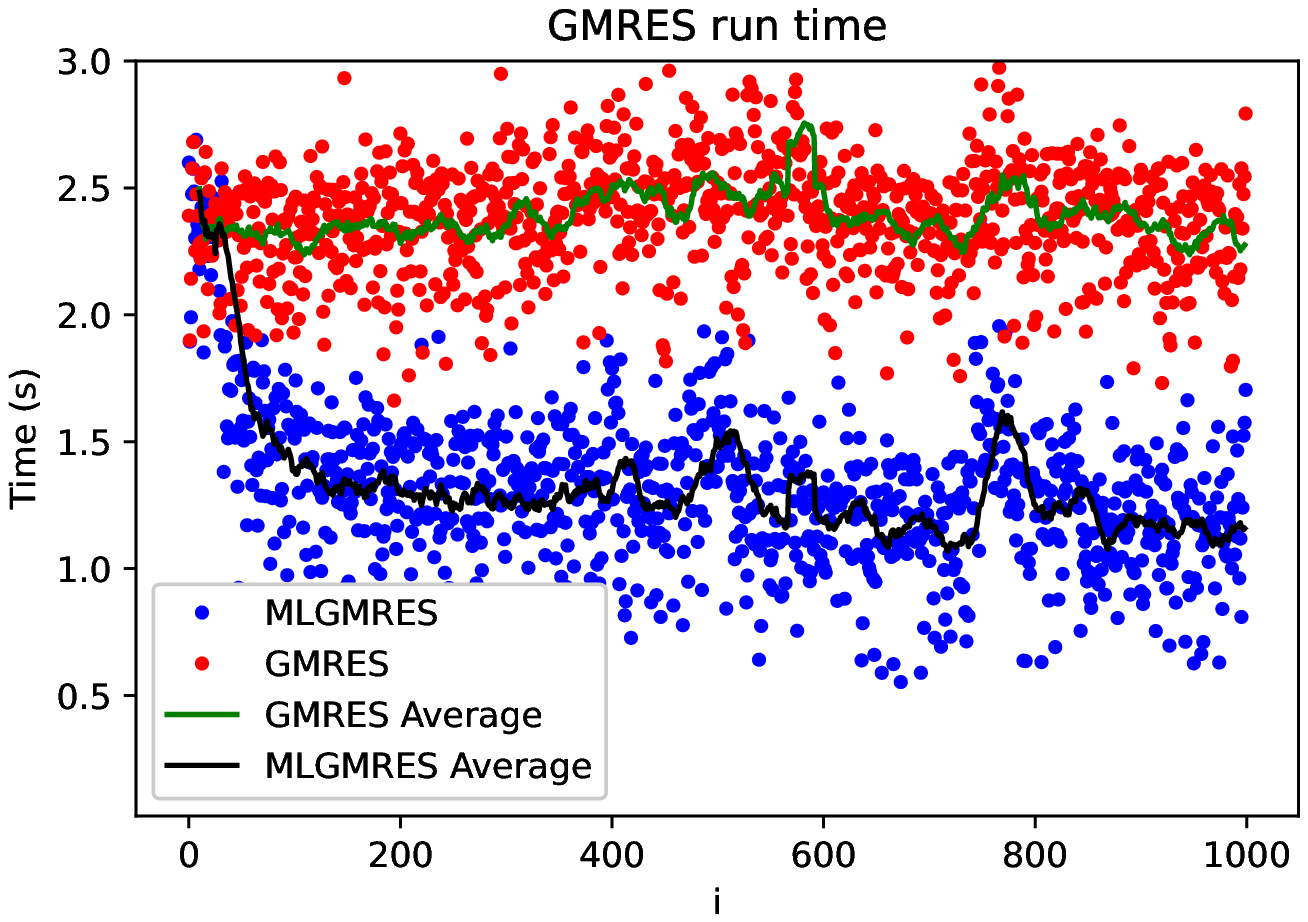}
    \includegraphics[width=0.49 \linewidth]{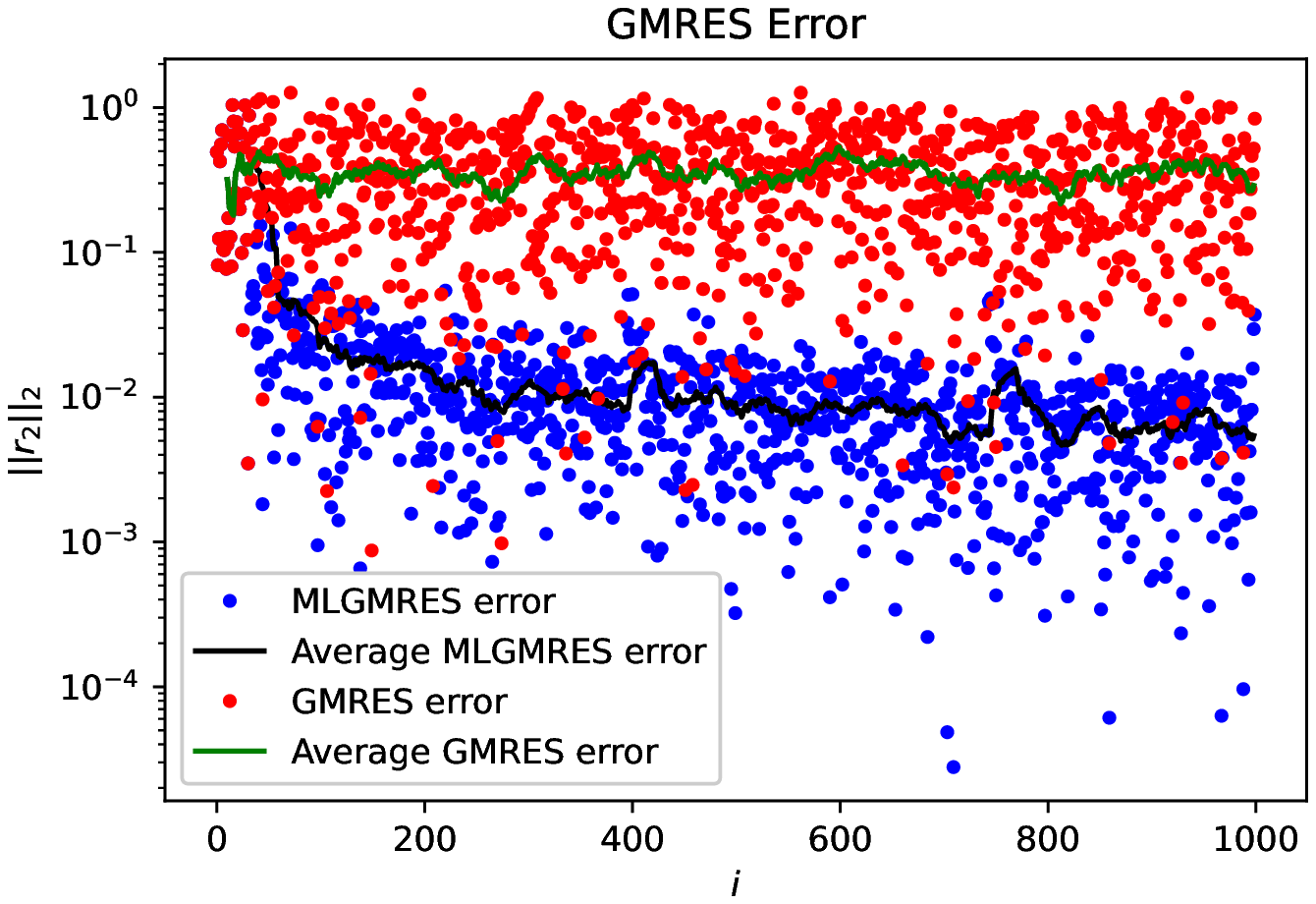}
    \caption{
        \emph{Left:} Speed up attained by MLGMRES (with adaptive FluidNet architecture) during the simulation. Tests ran on a shared compute node. Performance variation in the un-accelerated code (red points, green line) are due to background workloads on the test system.
        \emph{Right:} Error of GMRES at the 5th iteration during simulation.
    }
    \label{fig:FN1}
\end{figure}

\begin{figure}
    \centering
    \includegraphics[width=0.49 \linewidth]{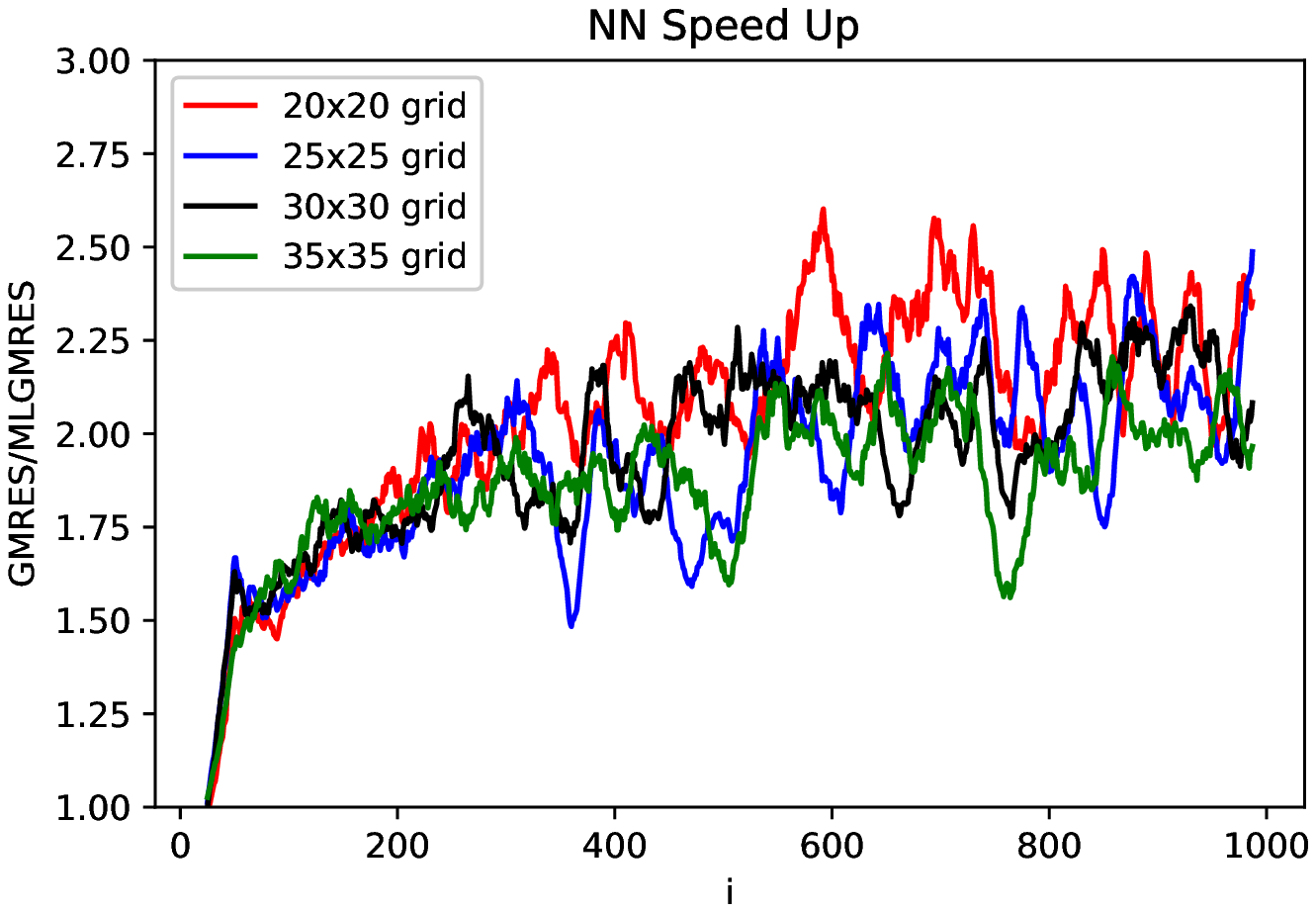}
    \includegraphics[width=0.49 \linewidth]{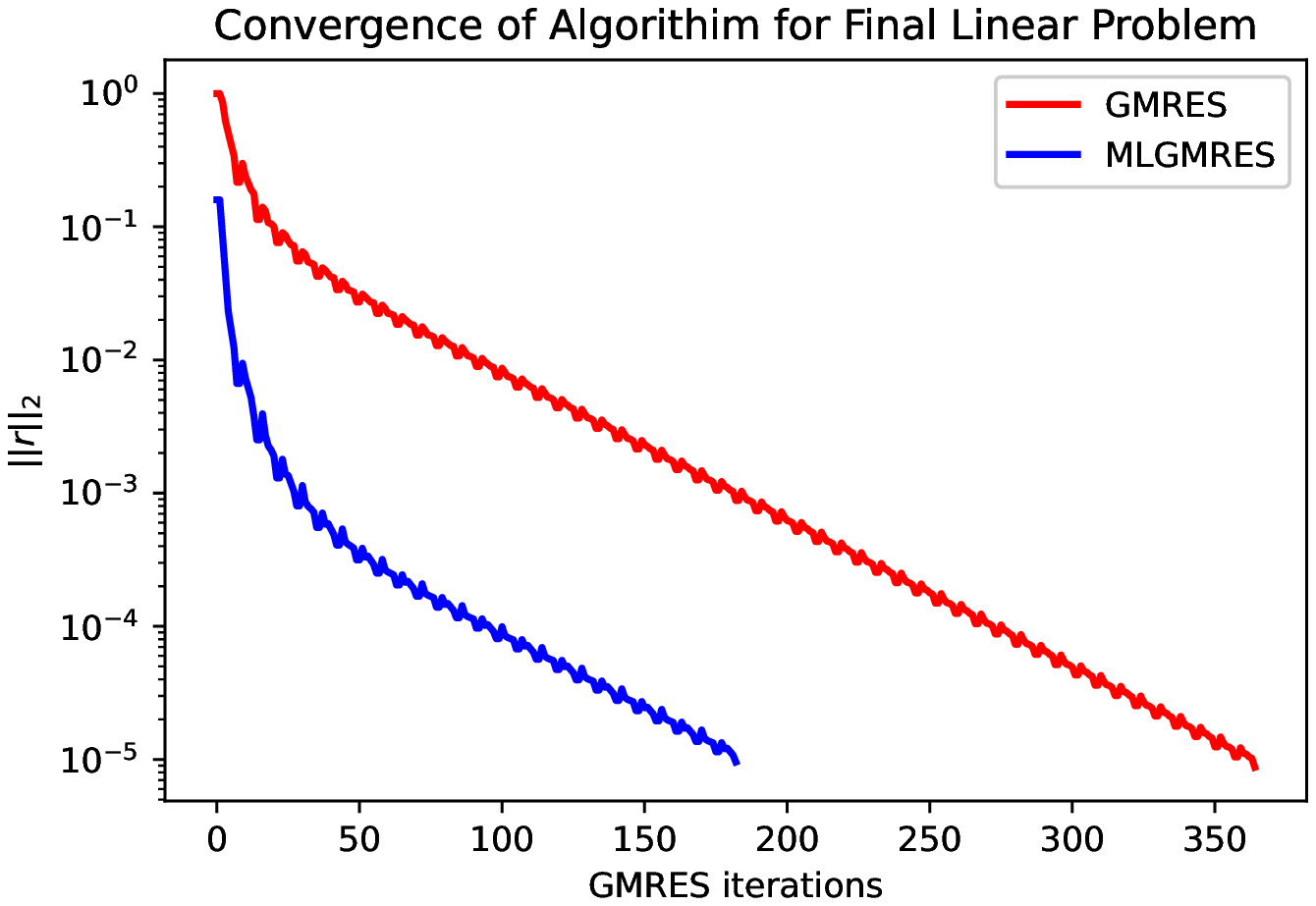}
    \caption{
        \emph{Left:} Speed up attained by MLGMRES (with adaptive FluidNet architecture) during the simulation
        \emph{Right:} Improvement in convergence rate of MLGMRES (with adaptive FluidNet architecture).
    }
        \label{fig:FN2}
\end{figure}

While purely convolutional  neural networks worked for a fairly simple RHS,  they do not work quite as
well on problems with multiple scales and different features. As a stepping-stone, we go from the ``monopole'' source term shown in Fig.~\ref{fig:LinearModel} to ``dipoles'' shown in Fig.~\ref{fig:DipolesExample}. For testing purposes these are not symmetric dipoles (in order to simulate systems with multiple scales). We find that purely-convolutional network architectures require more data to learn the relevant non-local relationships. To facilitate learning in these more complex scenarios, we implemented the
neural network of Section \ref{Secion_NetArch}. 
As test cases, we applied our AI-accelerator to
a sequence of dipoles with the scales for each pole randomly shifted (both on the domain and relative to each-other)
with different random overall scales for every single RHS. Figure \ref{fig:DipolesExample} demonstrates some of the many possibilities 
seen over the course of the simulation. 
In Fig. \ref{fig:FN1} we see that we are able to attain a significant improvement in the GMRES time-to-solution and in the initial residual. 
In fact we see about an order of magnitude improvement in the fifth iteration error of GMRES (left plot in Fig. \ref{fig:FN2}). This is particularly impressive given that this sequence of problems is significantly more complex than the previous example. On the right in Fig. \ref{fig:FN2} we see the GMRES error at each iteration for the final linear problem in the simulation.  As before, we see that the acceleration of the solver is mostly influenced by the improved rate of convergence rather than the initial residual (which is still relatively large compared to the target tolerance).

Finally, we tested the AI-accelerator for different resolutions. Since this network architecture inherently encodes non-locality, we use the same network for each resolution. Naturally, this means that the number of model parameters is fixed while we scale the the resolution up. These results are shown on the left in Fig. \ref{fig:FN2}. We see that as we increase the resolution, we get the same acceleration. So we see that even for difficult problems our network does not need to be overly complex (i.e. use an increasingly larger number of model parameters) for larger system sizes in order to accelerate the simulation. Consequently we are able to meet  our efficiency targets (relatively small number of parameters and trains quickly) with this network.

\section{Conclusion}

We have developed an \emph{in situ} real-time machine learning framework that
is capable of accelerating the time-to-solution of the GMRES($k$) algorithm. As
demonstrated by the application in Section~\ref{sec:prototype}, our real-time
algorithm can accelerate an iterative linear solver for the discrete Poisson equation.
With relatively little
input data (on average about 1 in 10 time steps yielded data that was
retained by Algorithm~\ref{al:luna}), and irrespective of problem size
(relative to number of network parameters) we are able to achieve a consistent
2$\times$ speedup. These results are promising for speed-up in
future applications. Our approach is flexible in that the neural network model
can be constructed from scratch and does not need to be pre-trained on large
amounts of expensive simulation data to get a performance boost. The approach
presented here only needs the desired iterative linear solver and our wrapper
augments the solver calls as outlined in Fig.
\ref{fig:chart} with the user's deep learning framework of choice. Beyond this,
the methodology presented here requires no further input from the user as the
simulation progresses.


\section{Acknowledgements}

This work was supported by the U.S. Department of Energy Office of Science, and
the Computing Sciences Summer Student program at Berkeley Lab. This research
used resources of the National Energy Research Scientific Computing Center
(NERSC), a U.S. Department of Energy Office of Science User Facility operated
under Contract No. DE-AC02-05CH11231. The authors are grateful to Stephen Whitelam, Daniel Herde, Steven Reeves, Deborah Bard, and Wibe A. de Jong for fruitful discussions.

\newpage

\bibliographystyle{hunsrt}


\end{document}